\documentclass[12pt]{article}
\usepackage[a4paper, total={6in, 8in}]{geometry}
\usepackage{graphicx, amsmath, bbold, tensor,xcolor} 
\usepackage{amsfonts}

\usepackage{verbatim}
\usepackage{amssymb}
\usepackage{dynkin-diagrams, braket}
\usepackage{tikz}
\usetikzlibrary{calc, arrows.meta,shapes,positioning}
\usepackage[sorting=none,style=numeric,citestyle=numeric-comp,backend=biber,doi=false,isbn=false,url=false,eprint=false,maxcitenames = 6,
mincitenames = 1,
uniquename = true,
uniquelist = true,
maxbibnames = 99]{biblatex}
\usepackage{caption}
\usepackage[T1]{fontenc}

\DeclareFieldFormat{pages}{#1}
\renewbibmacro{in:}{}
\usepackage[colorlinks, allcolors=blue]{hyperref}
\bibliography{bibliography} 
\DeclareFieldFormat[article, inproceedings]{title}{}

\title{Unification of Conformal Gravity and Internal Interactions}
\date{November 2023}
\author{Author}

\begin{document}

\author{Danai Roumelioti$^1$, Stelios Stefas$^1$, George Zoupanos$^{1,2,3}$}\date{}
\maketitle

\begin{center}
\itshape$^1$Physics Department, National Technical University, Athens, Greece\\
\itshape$^2$ Max-Planck Institut f\"ur Physik, M\"unchen, Germany\\
\itshape$^3$ Institut f\"ur Theoretische Physik der Universit\"at Heidelberg, Germany
\end{center}

\begin{center}
\emph{E-mails: \href{mailto:danai\_roumelioti@mail.ntua.gr}{danai\_roumelioti@mail.ntua.gr}, \href{mailto:dstefas@mail.ntua.gr}{dstefas@mail.ntua.gr}, \href{mailto:George.Zoupanos@cern.ch}{George.Zoupanos@cern.ch}}
\end{center}

\begin{abstract}
\noindent
Based on the observation that the dimension of the tangent space is not necessarily equal to the dimension of the corresponding curved manifold and on the known fact that gravitational theories can be formulated in a gauge theoretic way, we discuss how to describe all known interactions in a unified manner. This is achieved by enlarging the tangent group of the four-dimensional manifold to $SO(2,16)$, which permits the inclusion of both gauge groups, the one that describes gravity as a gauge theory as well as the $SO(10)$ describing the internal interactions. Moreover it permits the use of both Weyl and Majorana conditions imposed on the fermions, as to avoid the duplication of fermion multiplets of $SO(10)$ appearing in previous attempts. The gravity theory discussed in the present work is the Conformal Gravity which, after a spontaneous symmetry breaking, can lead either to Weyl Gravity or to the usual Einstein Gravity.
\end{abstract}

\section{Introduction}
An ultimate ambition of many theoretical physicists is the existence of a unification picture in which all fundamental interactions are included. An enormous amount of serious research activity has been carried out over decades that elaborated even the very appealing theoretical notion of extra dimensions, although the latter does not have any experimental support. Superstring theories \cite{Green2012-ul,polchinski_1998,Blumenhagen:2013fgp} have provided a solid framework of unification of all interactions; in particular the heterotic string theory \cite{GROSS1985253}, defined in ten dimensions, connected gravity in the form of supergravity, with the other interactions. More recently the AdS/CFT correspondence \cite{Maldacena:1997re} conjectured a relationship among Anti-de Sitter (AdS) spaces used in theories of quantum gravity (formulated in terms of string theory or M-theory) and conformal field theories (CFT) which are quantum field theories, including theories similar to the Yang–Mills ones that describe elementary particles. However, the successful description of elementary particles physics using gauge theories has not managed to be transferred to gravity so far. On the other hand, it is true that for more than a century, the General Theory of Relativity (GR) has been established as the theory describing gravitational interaction and has passed several tests over the years. In addition, it has been strongly emphasized that, contrary to the rest of the fundamental interactions, which are established in a gauge-theoretic way, GR is geometrically formulated, rendering gravity as an intrinsic property of the spacetime, related to its curvature. This discrepancy on the formulations of the theories of the fundamental interactions has been a major concern of theoretical physicists for several decades and it was not truly bypassed in the serious attempts mentioned above. Therefore, there might exist fruitful ground for attempts that attribute a direct, single description of all interactions based on gauge theories. In the present work we concentrate in the description of all interactions including gravity as gauge interactions. Moreover we provide a unified gauge theory which can accommodate gravity in its various forms with internal interactions described by the GUT $SO(10)$ \cite{FRITZSCH1975193, Georgi1999}.

The basic idea behind treating gravity as a gauge theory is not really surprising. Given that gravity is a diffeomorphism invariant theory, it is clearly invariant with respect to transformations whose parameters are functions of spacetime, just like in the local gauge theories and therefore, naturally, it has been long believed that GR can be formulated as a gauge theory \cite{PhysRev.101.1597,Kibble:1961ba,PhysRevLett.38.739}. The spin connection can be treated as the gauge field of the theory and would enter in the action using its field strength. The basic idea was used in a fundamental way in supergravity (see e.g. ref. \cite{freedman_vanproeyen_2012}) and recently it was transferred in the non-commutative gravity attempts too \cite{Chatzistavrakidis_2018,Manolakos:2019fle,Manolakos:2021rcl, Manolakos:2022ihc}. A more direct way along the line of unifying gravity as gauge theory with the rest fundamental interactions described as GUTs was suggested recently \cite{Percacci:1984ai,Percacci_1991,Nesti_2008,Nesti_2010,Krasnov:2017epi,Chamseddine2010,Chamseddine2016,Manolakos:2023hif,noncomtomos, Konitopoulos:2023wst}. It is based on the following observation: although usually the dimension of the tangent space is taken to be equal to the dimension of the corresponding curved manifold, the tangent group of a manifold of dimension d is not necessarily $SO(d)$ \cite{Weinberg:1984ke}. It is very interesting that one can consider higher than four dimensional tangent groups in a 4-dimensional spacetime which opens the possibility to achieve unification of gravity with internal interactions by gauging these higher-dimensional tangent groups.
Higher-derivative theories of gravity (see e.g. \cite{stelle.77, Stelle:1977ry, Alvarez-Gaume:2015rwa, hell2023degrees}, are known to be renormalizable \cite{stelle.77, Stelle:1977ry}. As such, these theories could play a role in the UV completion of gravity (see e.g. \cite{tHooft:2011aa, Adler:1982ri}). Weyl Gravity (WG) (see e.g. \cite{Maldacena:2011mk, Hell:2023rbf, Ghilencea:2023sti, Anastasiou:2016jix, Anastasiou:2020mik}) - a model whose action is made out of the square of the Weyl tensor - takes a predominant role among these theories and can result from Conformal Gravity (CG) by spontaneous symmetry breaking (SSB), as will be shown in section \ref{sec3}.

Dealing with high dimensional theories, there exist constraints that one has to take into account when aiming to construct realistic chiral theories for the internal interactions. These constraints are known from the construction of GUTs \cite{Georgi1999,FRITZSCH1975193,CHAPLINE1982461} and from theories with extra physical dimensions such as in the coset space dimensional reduction (CSDR) scheme \cite{forgacs,KAPETANAKIS19924,Kubyshin:1989vd,SCHERK197961,MANTON1981502,CHAPLINE1982461,LUST1985309,Manousselis_2004,Chatzistavrakidis:2009mh,Irges:2011de,Manolakos:2020cco,Patellis:2023npy}; similarly in the attempts to avoid doubling of the spectrum of chiral theories by imposing Majorana condition in the extra dimensions \cite{KAPETANAKIS19924,CHAPLINE1982461}. These constraints are recalled and used in section \ref{sec6}.

The structure of the present work is the following. In section \ref{sec2} we present in some detail the gauging of the AdS group, $SO(2,3)$, \cite{PhysRevD.21.1466,Ivanov:1980tw,Ivanov:1981wn,Kibble:1985sn} which leads after SSB to the Einstein Gravity (EG) with cosmological constant. In this way the gauging of the tangent group and its breaking are established, which are heavily used in the next sections. Section \ref{sec3} deals with the gauging of the conformal group, $SO(2,4)$, which leads to the CG \cite{Kaku:1978nz,Fradkin:1985am,Ivanov:1980tw,Chamseddine:2002fd}, while its breaking can lead, among others,  to either the WG or the EG. To our knowledge this is the first time that the breaking of the conformal group is done in a spontaneous way, i.e. without imposing constraints. The conformal group, its gauging and breakings are then used in the construction of the unified scheme based on $SO(2,16)$ of CG with internal interactions described by the GUT $SO(10)$ that will be discussed in section \ref{sec5}. Section \ref{sec6} will be devoted in the examination of fermions, and finally in section \ref{sec7} we will present our conclusions.

\section{Gauging of the Anti-de Sitter group \texorpdfstring{$SO(2, 3)$}{SO(3,2)}}\label{sec2}

When describing a $D-$dimensional theory of gravity as a $SO(p,q)$ gauge theory, a non-coordinate basis is being used (tetrad basis). In this case the spin connection and the vielbein are being used, instead of the metric. The first is regarded as the gauge field of the theory, while the latter relates vector fields from the non-coordinate tangent space basis back to the coordinate \cite{Witten:1988hc, Witten:1983}. The curvature tensor, comes naturally by its definition, as the field strength tensor (curvature two-form), and is only dependent on the spin connection. In the four-dimensional case the Einstein-Hilbert (E-H) Lagrangian, in this notation, is of the form

$$\sim \epsilon^{\mu \nu \rho \sigma} \epsilon_{a b c d} e_\mu{}^a e_\nu{}^b R_{\rho \sigma}{}^{cd},$$
where the $\mu,\nu,\rho,\sigma$ indices are running on the $SO(1,3)$ base manifold, while the $a,b,c,d$ are on the $SO(p,q)$ higher-dimensional tangent space. Such a Lagrangian cannot be directly retrieved by any gauge theory, and thus the E-H action in four dimensions does not originate from a gauge theory\footnote{In the three-dimensional case, the situation is different and the E-H action can be directly retrieved by a Chern-Simons action \cite{Witten:1988hc}.}. However, there exists a non-trivial way that E-H can be obtained by starting with a larger gauge symmetry action which subsequently breaks via SSB mechanism or by imposing constraints. The broken action may contain the desired E-H, along with other terms.

In this section we retrieve the E-H action as a spontaneously broken action. The natural choice of the larger group to be spontaneously broken would be that of the Poincar\'e group, $ISO(1,3)$. However, the distinct behaviour of the translations generator does not allow a full local gauge symmetry, as is the case of Lorentz transformations, thus another group is needed to be employed, in which all the generators would be on equal footing. The two candidate groups are the de Sitter (dS), $SO(1,4)$, and the Anti-de Sitter group (AdS), $SO(2,3)$. Both of these groups can be spontaneously broken to the Lorentz group, $SO(1,3)$, and both contain the same number of generators as the Poincar\'e group, but with the property that they can be denoted by a single gauge field, $\omega_\mu{}^{AB}$, where $A,B=1,\dots,5$. Here, following the approach of \cite{PhysRevD.21.1466}, and also \cite{Kibble:1985sn}, we are going to summarise the case of $SO(2,3)$, the AdS group. The case of $SO(1,4)$ has been presented in \cite{manolakosphd}.
The algebra of the $SO(2,3)$ group is the following:
\begin{equation}
    \left[J_{A B}, J_{C D}\right] = \eta_{BC}J_{AD}+\eta_{AD} J_{BC} -\eta_{AC} J_{BD} -\eta_{BD} J_{AC} ,
\end{equation}
where the metric of the gauge theory is $\eta_{AB}=\operatorname{diag}(-1,1,1,1,-1)$ with $A,B=1,\dots,5$, while the gauge connection for the gauge fields, $\omega_\mu{}^{AB}$, is $A_\mu = \frac{1}{2}\omega_\mu{}^{AB}J_{AB}$, where $J_{AB}$ are the ten generators of the AdS group. Employing the definition of the field strength tensor of $A_\mu$, 
\begin{equation}
F_{\mu \nu}=\left[D_\mu, D_\nu\right]=\partial_\mu A_\nu-\partial_\nu A_\mu+\left[A_\mu, A_\nu\right],
\end{equation}
and since $F_{\mu\nu} = \frac{1}{2}F_{\mu\nu}{}^{AB}J_{AB}$, we result with the expression
\begin{equation}
F_{\mu \nu}{}^{A B}=\partial_\mu \omega_\nu{}^{A B}-\partial_\nu \omega_\mu{}^{A B}+\omega_\mu{}^A{}_C \omega_\nu{}^{C B}-\omega_\nu{}^A{}_C \omega_\mu{}^{C B}.
\end{equation}
The only invariant that can be constructed that is also polynomial to the field strength tensor is the topological invariant that yields the Pontryagin index, 
\begin{equation}\label{initialaction}
S\sim \int d^4 x \epsilon^{\mu \nu \rho \sigma} F_{\mu \nu}{}^{A B} F_{\rho \sigma A B},
\end{equation} 
where $\epsilon^{\mu \nu \rho \sigma}$ is the Levi-Civita symbol. The above integral is a total divergence and also parity violating and C conserving (hence CP violating). Although it has been shown that there exists a nonpolynomial choice of Lagrangian \cite{West:1978nd}, in the works of \cite{PhysRevD.21.1466} and \cite{Kibble:1985sn} this choice was not eventually promoted due to the impractical expression of the integral. Instead, another choice is preferred which is also of second order in terms of the field strength tensor, but in this case in addition to the gauge field an auxiliary scalar field, $y^A$, is introduced, along with a dimensionful  parameter, $m$, satisfying a constraint equation, while contractions are being done by using the
tensor $\epsilon_{ABCDE}$. In this case the action is parity conserving, and is not a total divergence, as will be shown below, and takes the form:
\begin{equation}
S=a_{AdS}\int d^4 x\left(m y^E \epsilon_{A B C D E} \frac{1}{4}F_{\mu \nu}{}^{A B} F_{\rho \sigma}{}^{C D}\epsilon^{\mu \nu \rho \sigma} +\lambda\left(y^E y_E+m^{-2}\right)\right),
\end{equation}
where $a_{AdS}$ is a dimensionless coupling, and $\lambda$ is a parameter acting as a Lagrange multiplier, which imposes the constraint
\begin{equation}\label{gauge}
    y^Ey_E=-m^{-2}.
\end{equation}

Picking a specific gauge in which,
\begin{equation}
    y=y^0=(0,0,0,0,m^{-1}),
\end{equation}
the $SO(2,3)$ symmetry is spontaneously broken down to the little
group (isotropy subgroup) of $y^0$, i.e. the Lorentz group, $SO(1,3)$\footnote{The reader is reminded that in the initial $SO(2,3)$ symmetry metric, $\eta_{AB}$, the 1st and the 5th component where chosen to be the timelike components.}. Since $y^0$ is timelike, then $m^2>0$.

The action of the remaining symmetry, then becomes 
\begin{equation}
S_{\mathrm{SO}(1,3)}=\frac{a_{AdS}}{4}\int d^4 x \epsilon^{\mu \nu \rho \sigma} F_{\mu \nu}{}^{a b} F_{\rho \sigma}{}^{c d} \epsilon_{a b c d}.
\end{equation}
Defining the scaled gauge field, $e_\mu{}^a=m^{-1} \omega_\mu{}^{a 5}$, which corresponds to the five broken translation-like generators, we obtain from the components of the initial symmetry field strength tensor\footnote{The $SO(2,3)$ algebra decomposition in four-dimensional notation can be found in appendix \ref{apa1}.} 
\begin{equation}\label{tors}
F_{\mu \nu}{}^{a 5}=m (\partial_\mu e_\nu{}^a-\partial_\nu e_\mu{}^a-\omega_\mu{}^{a b} e_{\nu b}+\omega_\nu{}^{a b} e_{\mu b}),
\end{equation}
and
\begin{equation}\label{curv}
F_{\mu \nu}{}^{a b}=(\partial_\mu \omega_\nu{}^{a b}-\partial_\nu \omega_\mu{}^{a b}+\omega_\mu{}^{a}{}_{c} \omega_{\nu}{}^{cb}-\omega_\nu{}^{a}{}_{c} \omega_{\mu}{}^{cb})+m^2\left(e_\mu{}^a e_\nu{}^b-e_\nu{}^a e_\mu{}^b\right).
\end{equation}
The expression in the parenthesis in eq. \eqref{tors} is identified with the torsion tensor, $T_{\mu \nu}{}^a$, in the vielbein formalism description of GR, while the expression in the first parenthesis of eq. \eqref{curv} is the curvature two-form, $R_{\mu \nu}{}^{ab}$. The absence of a $F_{\mu \nu}{}^{a 5}$ term in the broken action, implies that the theory is torsion-free, since it can be set equal to zero. Substituting the curvature two-form in the broken action, one obtains
\begin{equation}
    \begin{aligned}
        S_{\mathrm{SO}(1,3)} &=\frac{a_{AdS}}{4}\int d^4 x \epsilon^{\mu \nu \rho \sigma} \epsilon_{a b c d}\left[R_{\mu \nu}{}^{a b}+m^2\left(e_\mu{}^a e_\nu{}^b-e_\mu{}^b e_\nu{}^a\right)\right]\\
        &\quad\qquad\qquad\qquad\times\left[R_{\rho \sigma}{}^{c d}+m^2\left(e_\rho{}^c e_\sigma{}^d-e_\rho{}^d e_\sigma{}^c\right)\right] \\
        &=\frac{a_{AdS}}{4}\int d^4 x \epsilon^{\mu \nu \rho \sigma} \epsilon_{a b c d}\left[R_{\mu \nu}{}^{a b}R_{\rho \sigma}{}^{c d}+4m^2 R_{\mu \nu}{}^{a b}e_\rho{}^c e_\sigma{}^d+4m^4e_\mu{}^a e_\nu{}^b e_\rho{}^c e_\sigma{}^d\right].
    \end{aligned}
\end{equation}

From the above expression it is obvious that the resulting action consists of three terms of the general form
\begin{equation}
    S_{\mathrm{SO}(1,3)}=\frac{a_{AdS}}{4}\int d^4 x \epsilon^{\mu \nu \rho \sigma} \epsilon_{a b c d}\left(\mathcal{L}_{R R}+ m^2 \mathcal{L}_{R e e}+ m^4\mathcal{L}_{eeee}\right),
\end{equation}
where $a_{AdS}>0$. The first term does not contribute to the field equations, as it yields the Gauss-Bonnet (G-B) topological invariant. The second term is the one identified with the E-H action, as it contains the Ricci scalar curvature, while the third term is a cosmological constant of order $m^4$, which causes the maximally symmetric solution of the field equations to be the AdS space\footnote{Recall that the expression of the E-H lagrangian in the presence of a cosmological constant is $\mathcal{L}_{\text{E-H}}\sim R-2\Lambda$, thus in our case the cosmological constant term is negative, as expected for the AdS space.},
\begin{equation}
F_{\mu \nu}{}^{a b}=0 \Rightarrow R_{\mu \nu}{}^{a b}=- m^2\left(e_\mu{}^a e_\nu{}^b-e_\nu{}^a e_\mu{}^b\right).
\end{equation}

Concluding, although the transformations of the gauge fields $e$ and $\omega$ can be obtained when choosing the Poincar\'e group as the gauge group, in order to result with an E-H equivalent action, one has to employ either the AdS or the dS group, starting from an action that is polynomial with respect to the field strength tensor, and include an auxiliary scalar field which satisfies a constraint and can be chosen to lie in a fixed gauge. Thus, a SSB is being induced and the original symmetry is being reduced to the Lorentz. The resulting action, although gauge invariant, is not a total divergence, contrary to \eqref{initialaction}, and despite the fact that it includes the G-B topological term, the rest of its terms can provide the usual equations of motion of EG in AdS space. In this way, we manage to describe the four-dimensional EG as a gauge theory.

\section{Gauging of the conformal group, \texorpdfstring{$SO(2,4)$}{SO(2,4)}}\label{sec3}

The conformal group, $SO(2,4)$, is the group of transformations on spacetime which leave invariant the null interval $ds^2 = \eta_{\mu \nu} dx^\mu dx^\nu = 0$. Field theories which are Lorentz invariant may or may not be scale invariant (e.g. the E-H action). By gauging the conformal group, and breaking it via the imposition of constraints method, it has been shown that Weyl's scale invariant theory of gravity can be retrieved \cite{Kaku:1978nz,Fradkin:1985am,Ivanov:1980tw,Chamseddine:2002fd}. In this work, for the first time, the breaking of the conformal gauge group is going to be spontaneous, induced by the introduction of a scalar field in the action, with the lagrange multiplier method.

In the first subsection we perform a SSB of the conformal group and result to the E-H action in the presence of a cosmological constant. In the second, we chose a different SSB, leading both to the E-H or Weyl action, depending on the relation between two of the gauge fields.

\subsection{Einstein-Hilbert and Weyl action from SSB of the conformal group in two steps}\label{31}

We wish to break the $SO(2,4)$ gauge symmetry via SSB mechanism. The point being contrary to the usual method, according to which the breaking of the original symmetry was done by imposition of constraints, we follow the more natural method of SSB introduced and used in section \ref{sec2}. Therefore the $SO(2,4)$ gauge group, whose algebra is isomorphic to those of $SU(4)$ and $SO(6)$, is going to be broken spontaneously to the $Sp_4 = SO(5)$ by introducing a scalar field in the vector representation (rep) of $SO(6)$, $6$, which takes vev in the $\langle 1\rangle$ component \cite{Slansky:1981yr} of the $6$ according to the branching rules of reps of $SO(6)$ to its maximal subgroup $SO(5)$:
\begin{equation}
\begin{aligned} 
    SO(6) &\supset SO(5) \\
    6 &=1+5.
\end{aligned}
\end{equation}
 As discussed in the previous section, \ref{sec2}, the $SO(2,3)$ breaks spontaneously to the $SO(1,3)$, when a scalar in the $5$ rep takes its vev in the $\langle 1,1\rangle$ component, and EG is finally obtained. Similarly, the following branching rule holds:  
\begin{equation}
\begin{aligned} 
    SO(5) &\supset SU(2)\times SU(2)\\
    5 &=(1,1)+(2,2),
\end{aligned}
\end{equation}
where the algebra of $SU(2)\times SU(2)$ is isomorphic to those of $SO(4)$ and $SO(1,3)$. A scalar in the vector rep $5$ of $SO(5)$ causes the final breaking that results in the algebra of the gauge group $SO(1,3)$, by taking vev in its $\langle 1,1\rangle$ component. Specifically, we start with the $SO(2,4)$ algebra of its generators:
\begin{equation}
    \left[J_{A B}, J_{C D}\right] = \eta_{BC}J_{AD}+\eta_{AD} J_{BC} -\eta_{AC} J_{BD} -\eta_{BD} J_{AC} ,
\end{equation}
where $A,B=1,\dots,6$ and $\eta_{AB}=(-1,1,1,1,-1,1)$. The reason of this choice of signature is for the two breakings planned convenience, since we are going to break at first a spacelike dimension (that will be represented by the 6th component of $\eta_{AB}$), while in the second step we are going to break a time-like (represented by the 5th component). A single gauge field, $A_{\mu}{}^{AB}$, can be denoted for all generators, $A_{\mu}=\frac{1}{2}A_{\mu}{}^{AB}J_{AB}$, thus the field strength tensor, $F_{\mu \nu}=\frac{1}{2}F_{\mu \nu}{}^{AB}J_{AB}$, is by definition
\begin{equation*}
    F_{\mu \nu}=[D_\mu , D_\nu]=\partial_\mu A_\nu -\partial_\nu A_\mu +[A_\mu ,A_\nu]\rightarrow
\end{equation*}
\begin{equation}
    F_{\mu \nu}{}^{AB}=\partial_\mu A_\nu{}^{AB} -\partial_\nu A_\mu{}^{AB} +A_\mu{}^{A}{}_{C}A_\nu{}^{CB}-A_\nu{}^{A}{}_{C}A_\mu{}^{CB}.
\end{equation}
Following a similar procedure to the one presented in section \ref{sec2}, we now construct an $SO(2,4)$ invariant quadratic action by introducing two scalar fields, $\phi^E$ and $\chi^F$, that belong to the vector rep $6$ of $SO(2,4)$, along with two dimensionful parameters, $m_\phi$ and $m_\chi$:
\begin{equation}
\begin{aligned}
    S_{SO(2,4)}=a_{CG}\int d^4x [&\epsilon^{\mu \nu \rho \sigma}\epsilon_{ABCDEF}\phi^E \chi^F m_\phi m_\chi \frac{1}{4} F_{\mu \nu}{}^{AB}F_{\rho \sigma}{}^{CD}+\\
    &+\lambda_\phi (\phi^E \phi_E -{m_\phi}^{-2})+\lambda_\chi (\chi^F \chi_F +{m_\chi}^{-2})], 
\end{aligned}
\end{equation}
where $a_{CG}$ a dimensionless coupling, and $ m_\phi\geq m_\chi$. The scalar field, $\phi^E$, fixed in the gauge that induces the desired breaking, is
\begin{equation}
    \phi^E=\phi^0=(0,0,0,0,0,{m_\phi}^{-1}),
\end{equation}
with ${m_\phi}^2>0$, since $\phi^0$ is spacelike. In this specific gauge, the component gauge field corresponding to the $5$ broken generators becomes:
\begin{equation}
    A_\mu{}^{j6}=m_\phi f_\mu{}^j,
\end{equation}
where $f$ is the scaled gauge field (in analogy to the scaled vielbein in the previous section)\footnote{Note that since $A_\mu{}^{AB}$ is antisymmetric under the permutation of its last two indices, then $f_\mu{}^6\sim A_\mu{}^{66}=0$. See also appendix \ref{apa2}.}.

Accordingly, the field strength tensor will be
\begin{equation}
    F_{\mu \nu}{}^{jk}=\partial_\mu A_\nu{}^{jk} -\partial_\nu A_\mu{}^{jk} +A_\mu{}^{j}{}_{l}A_\nu{}^{lk}-A_\nu{}^{j}{}_{l}A_\mu{}^{lk}-{m_\phi}^2(f_\mu{}^j f_\nu{}^k-f_\mu{}^kf_\nu{}^j),
\end{equation}
where $j,k=1,\dots,5$, and thus the broken, $SO(2,3)$-invariant, action takes the form:
\begin{equation}
    S_{SO(2,3)}=a_{CG}\int d^4x [\epsilon^{\mu \nu \rho \sigma}\epsilon_{ijklm} m_\chi \chi^m \frac{1}{4}F_{\mu \nu}{}^{ij}F_{\rho \sigma}{}^{kl} + \lambda_\chi (\chi^m\chi_m +{m_{\chi}^{-2}})]. 
\end{equation}
We now proceed to the second SSB by gauge fixing the second scalar field. The scalar field, this time belonging to the vector rep, $5$, of $SO(2,3)$, is being fixed in the gauge:
\begin{equation}
    \chi^m=\chi^0=(0,0,0,0,{m_\chi}^{-1}),
\end{equation}
where ${m_\chi}^2>0$, since $\chi^0$ is timelike.

At this point it is useful to be reminded that the conformal group, $SO(2,4)$, consists of $15$ generators, which in their usual $4-$dimensional rep\footnote{See appendix \ref{apa2}.} can be interpreted as: six Lorentz rotations, $M_{ab}$, four translations, $P_a$, four conformal boosts, $K_a$, and the dilatation generator, $D$. The corresponding gauge fields of these generators in respective order are denoted by: $\omega_\mu{}^{ab}, e_\mu{}^a,b_\mu{}^a,\tilde{a}_\mu$. On the other hand, the AdS group, $SO(2,3)$, consists of $10$ generators: $6$ Lorentz rotations, $M_{ab}$, and $4$ translations, $P_a$. 

Hence, by the breaking $SO(2,4)\rightarrow SO(2,3)$ the five generators that have been broken can be related to linear combinations of $P_a, K_a$ and the $D$. 

The gauge field $f_\mu{}^j$ in four-dimensional notation, according to the identifications of the generators in appendix \ref{apa2} becomes
\begin{equation}
    f_\mu{}^j=\big(b_\mu{}^a-e_\mu{}^a,-\tilde{a}_\mu\big),
\end{equation}
where $j=1,\dots ,5$, and $a,b=1,\dots,4$. Then, by the second breaking, which is related to the 5th component we also obtain
\begin{equation}
    A_\mu{}^{a5}=-{m_\chi}(b_\mu{}^a+e_\mu{}^a).
\end{equation}
Thus, $4$ more generators are broken, and now only the $6$ Lorentz rotations generators, $M_{ab}$, remain unbroken. Gathering all the scaled gauge fields, we find that
\begin{align}
       A_\mu{}^{j6}=m_\phi f_\mu{}^j \Rightarrow \begin{cases}
           A_\mu{}^{a6}={m_\phi}(b_\mu{}^a-e_\mu{}^a) \\
           A_\mu{}^{56}=-m_\phi \tilde{a}_\mu \\
       \end{cases} 
       \text{and}\quad
     A_\mu{}^{a5}=-{m_\chi} (b_\mu{}^a+e_\mu{}^a),
\end{align}
while the remaining, unbroken gauge fields correspond to the spin connection,
\begin{equation}
    A_\mu{}^{ab}=\omega_\mu{}^{ab}.
\end{equation}
The substitution of the above in the field strength tensor leads to the following expression:
\begin{align}\label{curv2}
F_{\mu \nu}{}^{a b}= &\partial_\mu \omega_\nu{}^{a b}-\partial_\nu \omega_\mu{}^{a b}-\omega_\mu{}^{a c} \omega_{\nu c}{}^b+\omega_\nu{}^{a c} \omega_{\mu c}{}^b \nonumber\\
&+( {m_\chi}^2-{m_\phi}^2)\left(e_\mu{}^a e_\nu{}^b-e_\nu{}^a e_\mu{}^b+b_\mu{}^a b_\nu{}^b-b_\nu{}^a b_\mu{}^b\right)\nonumber\\
&-( {m_\chi}^2+{m_\phi}^2)(b_\mu{}^a e_\nu{}^b-b_\nu{}^a e_\mu{}^b+e_\mu{}^a b_\nu{}^b-e_\nu{}^a b_\mu{}^b)\longrightarrow\nonumber\\
F_{\mu \nu}{}^{a b}=&R_{\mu \nu}{}^{a b}+( {m_\chi}^2-{m_\phi}^2)\left(e_\mu{}^a e_\nu{}^b-e_\nu{}^a e_\mu{}^b+b_\mu{}^a b_\nu{}^b-b_\nu{}^a b_\mu{}^b\right)\\
&-( {m_\chi}^2+{m_\phi}^2)(b_\mu{}^a e_\nu{}^b-b_\nu{}^a e_\mu{}^b+e_\mu{}^a b_\nu{}^b-e_\nu{}^a b_\mu{}^b).\nonumber
\end{align}

Before reaching the final form of the broken action, it is worth noting that (like in the previous section), the $F_{\mu \nu}{}^{a5}$ contains now both the torsion tensors with respect to the gauge fields $e$ and $b$:
\begin{align}
\label{tors2}
F_{\mu \nu}{}^{a 5}&=-{m_\chi} (\partial_\mu e_\nu{}^a-\partial_\nu e_\mu{}^a-\omega_\mu{}^{a b} e_{\nu b}+\omega_\nu{}^{a b} e_{\mu b}+\partial_\mu b_\nu{}^a-\partial_\nu b_\mu{}^a-\omega_\mu{}^{a b} b_{\nu b}+\omega_\nu{}^{a b} b_{\mu b}) + \nonumber \\
&+{m_\phi}^2[\tilde{a}_\mu( e_\nu{}^a -b_\nu{}^a)-\tilde{a}_\nu( e_\mu{}^a-b_\mu{}^a)]\longrightarrow \nonumber\\
F_{\mu \nu}{}^{a 5}&=-{m_\chi}[ T_{\mu \nu}{}^a(e)+T_{\mu \nu}{}^a(b)] +{m_\phi}^2[\tilde{a}_\mu( e_\nu{}^a -b_\nu{}^a)-\tilde{a}_\nu( e_\mu{}^a-b_\mu{}^a)].
\end{align}
The final action, being quadratic in terms of the field strength tensor, \eqref{curv2}, does not include in any way the $F_{\mu \nu}{}^{a 5}$ field. This suggests that we may set  $F_{\mu \nu}{}^{a 5}$ equal to zero, which in turn means that we can also set $\tilde{a}_\mu=0$, and the theory might result being torsion-free.

Finally, given the above remarks, the final action of the theory, after the two successive SSBs, results being
\begin{equation}
\begin{aligned}\label{act2}
    S_{SO(1,3)} =\frac{a_{CG}}{4}\int &d^4x \epsilon^{\mu \nu \rho \sigma}\epsilon_{abcd}F_{\mu \nu}{ }^{a b} F_{\rho \sigma}{ }^{c d}= \\
    =\frac{a_{CG}}{4}\int & d^4x \epsilon^{\mu \nu \rho \sigma}\epsilon_{abcd}[R_{\mu \nu}{ }^{a b} R_{\rho \sigma}{ }^{c d} +\\
    &+2( {m_\chi}^2-{m_\phi}^2)R_{\mu \nu}{ }^{a b}(e_\rho{}^c e_\sigma{}^d-e_\sigma{}^c e_\rho{}^d+b_\rho{}^c b_\sigma{}^d-b_\sigma{}^c b_\rho{}^d)\\
    &-2( {m_\chi}^2+{m_\phi}^2)R_{\mu \nu}{ }^{a b}(b_\rho{}^c e_\sigma{}^d-b_\sigma{}^c e_\rho{}^d+e_\rho{}^c b_\sigma{}^d-e_\sigma{}^c b_\rho{}^d)\\
 &-2( {m_\chi}^4-{m_\phi}^4)\left(e_\mu{}^a e_\nu{}^b-e_\nu{}^a e_\mu{}^b+b_\mu{}^a b_\nu{}^b-b_\nu{}^a b_\mu{}^b\right) \\
&\qquad\qquad\qquad\qquad\times(b_\rho{}^c e_\sigma{}^d-b_\sigma{}^c e_\rho{}^d+e_\rho{}^c b_\sigma{}^d-e_\sigma{}^c b_\rho{}^d)\\
    &+( {m_\chi}^2-{m_\phi}^2)^2\left(e_\mu{}^a e_\nu{}^b-e_\nu{}^a e_\mu{}^b+b_\mu{}^a b_\nu{}^b-b_\nu{}^a b_\mu{}^b\right)\\
    &\qquad\qquad\qquad\qquad\times(e_\rho{}^c e_\sigma{}^d-e_\sigma{}^c e_\rho{}^d+b_\rho{}^c b_\sigma{}^d-b_\sigma{}^c b_\rho{}^d)\\
    & +( {m_\chi}^2+{m_\phi}^2)^2\left(b_\mu{}^a e_\nu{}^b-b_\nu{}^a e_\mu{}^b+e_\mu{}^a b_\nu{}^b-e_\nu{}^a b_\mu{}^b\right)\\
    &\qquad\qquad\qquad\qquad\times(b_\rho{}^c e_\sigma{}^d-b_\sigma{}^c e_\rho{}^d+e_\rho{}^c b_\sigma{}^d-e_\sigma{}^c b_\rho{}^d)].
\end{aligned}
\end{equation}
The $F_{\mu \nu}{}^{56}$ component of the field strength tensor is also absent from the action, thus we may set,
\begin{equation}
\begin{aligned}
    F_{\mu \nu}{}^{56}={m_\phi}[\partial_\mu \tilde{a}_\nu -\partial_\nu \tilde{a}_\mu-m_\chi(e_{\mu a}b_\nu{}^a-e_{\nu a}b_\mu{}^a)]=0.
\end{aligned}    
\end{equation}
The above, since $\tilde{a}_\mu=0$, leads to the following relation among $e$ and $b$:
\begin{equation}\label{efa}
    e_{\mu a}b_\nu{}^a-e_{\nu a}b_\mu{}^a=0.
\end{equation}

\subsubsection{When $b_\mu{}^a=ae_\mu{}^a$ - Einstein-Hilbert action in the presence of a cosmological constant}\label{casea}
As a solution of eq. \eqref{efa}, we choose that of $b$ and $e$ being proportional. Setting $b_\mu{}^a=ae_\mu{}^a$, as proposed in \cite{Chamseddine:2002fd}, then the final action becomes
\begin{equation}
\begin{aligned}
\label{34}
 S_{SO(1,3)} =\frac{a_{CG}}{4}\int d^4x &\epsilon^{\mu \nu \rho \sigma}\epsilon_{abcd}F_{\mu \nu}{ }^{a b} F_{\rho \sigma}{ }^{c d}= \\
    =\frac{a_{CG}}{4}\int d^4x \epsilon^{\mu \nu \rho \sigma}\epsilon_{abcd} & \Big[R_{\mu \nu}{ }^{a b} R_{\rho \sigma}{ }^{c d} +4\Big({m_\chi}^2(1-a)^2-{m_\phi}^2(1+a)^2\Big)R_{\mu \nu}{ }^{a b}e_\rho{}^c e_\sigma{}^d\\
    & +4\Big({m_\chi}^2(1-a)^2-{m_\phi}^2(1+a)^2\Big)^2 e_\mu{}^a e_\nu{}^b e_\rho{}^c e_\sigma{}^d\Big]
\end{aligned}
\end{equation}

Similarly to the previous section, \ref{sec2}, the first term of the action \eqref{34} is a G-B topological term, the second term is the E-H action equivalent and the third is the cosmological constant. From the above action when ${m_\chi}^2/{m_\phi}^2>(1+a)^2/(1-a)^2$ GR in AdS space is retrieved\footnote{It is worth mentioning that one can obtain exactly the same results by using an antisymmetric tensor rep for the scalar field, instead of two vector reps for two scalars, as shown in appendix \ref{apb}.}. 

In the case where $m_\phi=m_\chi\equiv m$, we obtain the action:
\begin{equation}
\begin{aligned}
\label{35}
  S_{SO(1,3)} =\frac{a_{CG}}{4}\int &d^4x \epsilon^{\mu \nu \rho \sigma}\epsilon_{abcd}F_{\mu \nu}{ }^{a b} F_{\rho \sigma}{ }^{c d} =\\
    =\frac{a_{CG}}{4}\int & d^4x \epsilon^{\mu \nu \rho \sigma}\epsilon_{abcd}\Big[R_{\mu \nu}{ }^{a b} R_{\rho \sigma}{ }^{c d} -16{m}^2aR_{\mu \nu}{ }^{a b} e_\rho{}^c e_\sigma{}^d
    +64{m}^4 a^2e_\mu{}^a e_\nu{}^b e_\rho{}^c e_\sigma{}^d\Big],
\end{aligned}
\end{equation}
which, for $a<0$ describes GR in AdS space.

\subsubsection{When $b_\mu{}^a=-\frac{1}{4}(R_\mu{}^a+\frac{1}{6}R e_\mu{}^a)$ and $m_\phi=m_\chi$ - Weyl action}
\label{caseb}

This relation among $b$ and $e$, which is again solution of \eqref{efa}, was suggested in refs \cite{Kaku:1978nz} and \cite{freedman_vanproeyen_2012}. Choosing this relation among $e$ and $b$, we obtain the following action:
\begin{equation}
    \begin{aligned}
             S=\frac{a_{CG}}{4}\int d^4 x \epsilon^{\mu \nu \rho \sigma} \epsilon_{a b c d}&\left[R_{\mu \nu}{}^{a b}+\frac{1}{2}\left(m e_\mu{}^{[a} R_\nu{}^{b]}-me_\nu{}^{[a} R_\mu{}^{b]}\right)-\frac{1}{3} m^2 R e_\mu{}^{[a} e_\nu{}^{b]}\right]\\
             &\left[R_{\rho \sigma}{}^{c d}+\frac{1}{2}\left(m e_\rho{}^{[c} R_\sigma{}^{d]}-m e_\sigma{}^{[c} R_\rho{}^{d]}\right)-\frac{1}{3} m^2 R e_\rho{}^{[c} e_\sigma{}^{d]}\right],
    \end{aligned}
\end{equation}
where $m \equiv m_\phi=m_\chi$. Taking into account that now the gauge field is the rescaled vierbein, $\tilde{e}_\mu{}^{a}=m e_\mu{}^{a}$, and recalling that $R_{\mu \nu}{}^{a b}=-R_{\nu \mu}{}^{a b}$, we get
\begin{equation}
    \begin{aligned}
             S=\frac{a_{CG}}{4}\int d^4 x \epsilon^{\mu \nu \rho \sigma} \epsilon_{a b c d}&\left[R_{\mu \nu}{}^{a b}-\frac{1}{2}\left(\tilde{e}_\mu{}^{[a} R_\nu{}^{b]}-\tilde{e}_\nu{}^{[a} R_\mu{}^{b]}\right)+\frac{1}{3} R \tilde{e}_\mu{}^{[a} \tilde{e}_\nu{}^{b]}\right]\\
             &\left[R_{\rho \sigma}{}^{c d}-\frac{1}{2}\left( \tilde{e}_\rho{}^{[c} R_\sigma{}^{d]}- \tilde{e}_\sigma{}^{[c} R_\rho{}^{d]}\right)+\frac{1}{3} R \tilde{e}_\rho{}^{[c} \tilde{e}_\sigma{}^{d]}\right].
    \end{aligned}
\end{equation}
We notice that the above action is equal to 
\begin{align}
             S=\frac{a_{CG}}{4}\int d^4 x \epsilon^{\mu \nu \rho \sigma} \epsilon_{a b c d}C_{\mu \nu}{}^{a b}C_{\rho \sigma}{}^{c d},
\end{align}
where $C_{\mu \nu}{}^{a b}$ is the Weyl conformal tensor. This action leads to the well-know action,
\begin{equation}\label{weyll}
S =2 a_{CG}\int \mathrm{d}^4 x\left(R_{\mu \nu} R^{\nu \mu}-\frac{1}{3} R^2\right),
\end{equation}
which describes the four-dimensional scale invariant Weyl theory of gravity.

\subsection{Einstein-Hilbert and Weyl action from SSB of the conformal group, by using a scalar in the adjoint representation}
\label{AdjointSubsection}

In the past, in order to construct the four-dimensional CG, one had to start by gauging the conformal group, $SO(2,4)$, and impose constraints in order to retrieve WG \cite{Kaku:1978nz, Fradkin:1985am}. Here, instead, we use SSB mechanism, and the $SO(2,4)$ gauge group is being reduced directly to the $SO(1,3)$ by a scalar field belonging to the adjoint rep. This can lead either to the four-dimensional EG or to WG.

The gauge group, $SO(2,4)$, as mentioned in the previous subsection, comprises of fifteen generators. Those generators in four-dimensional notation consist of six Lorentz transformations, $M_{ab}$, four translations, $P_a$, four special conformal transformations (conformal boosts), $K_a$, and the dilatation, $D$.

The gauge connection, $A_\mu$, as an element of the $SO(2,4)$ algebra, can be expanded in terms of the generators as
\begin{equation}
A_\mu= \frac{1}{2}\omega_\mu{}^{a b} M_{a b}+e_\mu{}^a P_a+b_\mu{}^a K_a+\tilde{a}_\mu D,
\end{equation}
where, for each generator a gauge field has been introduced\footnote{The rep chosen, along with the commutation and anticommutation relations of the generators, can be found in appendix \ref{apa2}.}. The gauge field related to the translations is identified as the vierbein, while the one of the Lorentz transformations is identified as the spin connection. The field strength tensor is of the form
\begin{equation}\label{fst}
F_{\mu \nu}=\frac{1}{2}R_{\mu \nu}{}^{a b} M_{a b}+\tilde{R}_{\mu \nu}{}^a P_a+R_{\mu \nu}{}^a K_a+R_{\mu \nu} D,
\end{equation}
where
\begin{equation}\label{curves}
\begin{aligned}
R_{\mu \nu}{}^{a b} & =\partial_\mu \omega_\nu{}^{a b}-\partial_\nu \omega_\mu{}^{a b}-\omega_\mu{}^{a c} \omega_{\nu c}{}^b+\omega_\nu{}^{a c} \omega_{\mu c}{}^b-8 e_{[\mu}{}^{[a} b_{\nu]}{}^{b]} \\
& =R_{\mu \nu}^{(0) a b}-8 e_{[\mu}{}^{[a} b_{\nu]}{}^{b]}, \\
\tilde{R}_{\mu \nu}{}^a & =\partial_\mu e_\nu{}^a-\partial_\nu e_\mu{}^a+\omega_\mu{}^{a b} e_{\nu b}-\omega_\nu{}^{a b} e_{\mu b}-2 \tilde{a}_{[\mu} e_{\nu]}{}^a \\
& =T_{\mu \nu}^{(0) a}-2 \tilde{a}_{[\mu} e_{\nu]}{}^a, \\
R_{\mu \nu}{}^a & =\partial_\mu b_\nu{}^a-\partial_\nu b_\mu{}^a+\omega_\mu{}^{a b} b_{\nu b}-\omega_\nu{}^{a b} b_{\mu b}+2 \tilde{a}_{[\mu} b_{\nu]}{}^a\\
&=T_{\mu \nu}^{(0) a}(b)+2 \tilde{a}_{[\mu} b_{\nu]}{}^a,\\
R_{\mu \nu} & =\partial_\mu \tilde{a}_\nu-\partial_\nu \tilde{a}_\mu+4 e_{[\mu}{}^a b_{\nu] a},
\end{aligned}
\end{equation}
where $T_{\mu \nu}^{(0) a}$ and $R_{\mu \nu}^{(0) a b}$ are the torsion and curvature component tensors in the four-dimensional vierbein formalism of GR, like in the previous section, \ref{sec2}, while $T_{\mu \nu}^{(0) a}(b)$ is the torsion tensor related to the gauge field $b_\mu{}^a$. 

We shall start again by choosing the parity conserving action, which is quadratic in terms of the field strength tensor \eqref{fst}, in which we have introduced a scalar that belongs to the adjoint rep, $15$, of $SO(6) \sim SO(2,4)$ along with a dimensionful parameter, $m$:
\begin{equation}
    S_{SO(2,4)}=a_{CG}\int d^4x [\operatorname{tr} \epsilon^{\mu \nu \rho \sigma} m\phi F_{\mu \nu}F_{\rho \sigma}+(\phi^2-m^{-2} \mathbb{1}_4)], 
\end{equation}
where the trace is defined as $ \operatorname{tr}\rightarrow \epsilon_{abcd} [\text{Generators}]^{abcd}$.

The scalar expanded on the generators is:
\begin{equation}
\phi=\phi^{a b} M_{a b}+\tilde{\phi}^a P_a+\phi^a K_a+\tilde{\phi} D,
\end{equation}

In accordance with \cite{Li:1973mq}, we pick the specific gauge in which $\phi$ is diagonal of the form $\operatorname{diag}(1,1,-1,-1)$. Specifically we choose $\phi$ to be only in the direction of the dilatation generator $D$:
\begin{equation}
    \phi=\phi^0=\tilde{\phi}D \xrightarrow{\phi^2=m^{-2}\mathbb{1}_4}\phi=-2m^{-1} D.
\end{equation}
In this particular gauge the action reduces to
\begin{equation}
    S=-2a_{CG}\int d^4x \operatorname{tr} \epsilon^{\mu \nu \rho \sigma} F_{\mu \nu}F_{\rho \sigma}D, 
\end{equation}
and the gauge fields $e,b$ and $\tilde{a}$ become scaled as $me,mb$ and $m\tilde{a}$ correspondingly.
After straightforward calculations, using the expansion of the field strength tensor as in eq. \eqref{fst}, and the anticommutation relations of the generators, we obtain: 
\begin{equation}
\begin{gathered}
    S=-2a_{CG}\int d^4x \operatorname{tr} \epsilon^{\mu \nu \rho \sigma}\Big[\frac{1}{4}R_{\mu \nu}{}^{ab}R_{\rho \sigma}{}^{cd}M_{ab}M_{cd}D\\
    +i\epsilon_{abcd}(R_{\mu \nu}{}^{ab}R_{\rho \sigma}{}^{c} K^d D - R_{\mu \nu}{}^{ab}\tilde{R}_{\rho \sigma}{}^{c}P^{d}D)+(\frac{1}{2}\tilde{R}_{\mu \nu}{}^{a}R_{\rho \sigma} + 2\tilde{R}_{\mu \nu}{}^{a}R_{\rho \sigma}{}^{b})M_{ab}\\
    +(\frac{1}{4}R_{\mu \nu}R_{\rho \sigma}- 2\tilde{R}_{\mu \nu}{}^{a}R_{\rho \sigma a})D   
    \Big].
\end{gathered}
\end{equation}
In this point we employ the trace on the several generators and their products. In particular:
\begin{equation}
\begin{gathered}
   \operatorname{tr}[K^{d}D]=\operatorname{tr}[P^{d}D]=\operatorname{tr}[M_{ab}]= \operatorname{tr}[D]=0, \\
        \text{and}\quad \operatorname{tr}[M_{ab}M_{cd}D]=-\frac{1}{2}\epsilon_{abcd}.
\end{gathered}
\end{equation}
The resulting broken action is:
\begin{equation}
\label{BrokenActionConformal}
     S_{\mathrm{SO}(1,3)}=\frac{a_{CG}}{4}\int d^4x \epsilon^{\mu \nu \rho \sigma}\epsilon_{abcd}R_{\mu \nu}{}^{ab}R_{\rho \sigma}{}^{cd},
\end{equation}
while its invariance has obviously been reduced only to Lorentz. Before continuing, we notice that there is no term containing the field $\tilde{a}_\mu$ in any way present in the action. Thus, as in the previous subsection, \ref{31}, we may set $\tilde{a}_\mu=0$\footnote{ Let us note that since $\tilde{a}_\mu$ is the gauge field corresponding to the dilatation generator, $D$, by switching off $\tilde{a}_\mu$ or by making it heavy due to SSB, it remains a global symmetry corresponding to scale invariance. The latter is broken by the presence of dimensionful parameters as the cosmological constant.\label{footnote}}. This simplifies the form of the two component field strength tensors related to the $P$ and $K$ generators:
\begin{equation}
\begin{aligned}
  \tilde{R}_{\mu \nu}{}^a & =mT_{\mu \nu}^{(0) a}-2 m^2\tilde{a}_{[\mu} e_{\nu]}{}^a \longrightarrow mT_{\mu \nu}^{(0) a}, \\
R_{\mu \nu}{}^a &=mT_{\mu \nu}^{(0) a}(b)+2m^2 \tilde{a}_{[\mu} b_{\nu]}{}^a \longrightarrow mT_{\mu \nu}^{(0) a}(b).
\end{aligned}
\end{equation}
The absence of the above field strength tensors in the action, allows us to also set $\tilde{R}_{\mu \nu}{}^a=R_{\mu \nu}{}^a=0$, and thus to obtain a torsion-free theory. Since $R_{\mu \nu}$ is also absent from the expression of the broken action, it may also be set equal to zero. From its definition in eq. \eqref{curves}, then we obtain the following relation among $e$ and $b$: 
\begin{equation}\label{ef}
    e_\mu{}^a b_{\nu a}-e_{\nu}{}^{a}b_{\mu a}=0,
\end{equation}
which is an equivalent result to the eq. \eqref{efa}. As in \ref{31}, the above result reinforces one to consider solutions that relate $e$ and $b$.

Here we examine, as in \ref{casea} and \ref{caseb}, the same two possible solutions of eq. \eqref{ef}.

\subsubsection{When $b_\mu{}^a=ae_\mu{}^a$ - Einstein-Hilbert action in the presence of a cosmological constant}
\label{subsubsectionA}
In this case, first proposed in \cite{Chamseddine:2002fd}, by a simple substitution we obtain:
\begin{equation}
\begin{aligned}
        S_{\mathrm{SO}(1,3)} =\frac{a_{CG}}{4}\int d^4 x \epsilon^{\mu \nu \rho \sigma} \epsilon_{a b c d}&\left[R_{\mu \nu}^{(0) a b}-4m^2a\left(e_\mu{}^a e_\nu{}^b-e_\mu{}^b e_\nu{}^a\right)\right]\\
       &\left[R_{\rho \sigma}^{(0) c d}-4m^2a\left(e_\rho{}^c e_\sigma{}^d-e_\rho{}^d e_\sigma{}^c\right)\right]\longrightarrow\\
        S_{\mathrm{SO}(1,3)} =\frac{a_{CG}}{4}\int d^4 x \epsilon^{\mu \nu \rho \sigma} \epsilon_{a b c d}&[R_{\mu \nu}^{(0) a b}-8m^2ae_\mu{}^a e_\nu{}^b]\left[R_{\rho \sigma}^{(0) c d}-8m^2ae_\rho{}^c e_\sigma{}^d\right],
       \end{aligned}
\end{equation}
which, as in \ref{casea} in the two-step breaking case, yields
\begin{equation}\label{so24finalaction}
\begin{aligned}
             S_{\mathrm{SO}(1,3)}=\frac{a_{CG}}{4}\int d^4 x \epsilon^{\mu \nu \rho \sigma} \epsilon_{a b c d}[R_{\mu \nu}^{(0) a b}R_{\rho \sigma}^{(0) c d}-16m^2aR_{\mu \nu}^{(0) a b}e_\rho{}^c e_\sigma{}^d+\\
             +64m^4a^2 e_\mu{}^a e_\nu{}^b e_\rho{}^c e_\sigma{}^d].
\end{aligned}
\end{equation}

This action consists of three terms: one G-B topological term, the E-H action, and a cosmological constant, and is identical to the one obtained previously, in the two-step breaking, \eqref{35}, which for $a<0$ describes GR in AdS space.

\subsubsection{When $b_\mu{}^a=-\frac{1}{4}(R_\mu{}^a+\frac{1}{6}R e_\mu{}^a)$ - Weyl action}

This relation among $b$ and $e$, which is again solution of \eqref{ef}, was suggested in refs \cite{Kaku:1978nz} and \cite{freedman_vanproeyen_2012}. Taking this into account we obtain the following action:
\begin{equation}
    \begin{aligned}
             S=\frac{a_{CG}}{4}\int d^4 x \epsilon^{\mu \nu \rho \sigma} \epsilon_{a b c d}&\left[R_{\mu \nu}^{(0) a b}+\frac{1}{2}\left(m e_\mu{}^{[a} R_\nu{}^{b]}-me_\nu{}^{[a} R_\mu{}^{b]}\right)-\frac{1}{3} m^2 R e_\mu{}^{[a} e_\nu{}^{b]}\right]\\
             &\left[R_{\rho \sigma}^{(0) c d}+\frac{1}{2}\left(m e_\rho{}^{[c} R_\sigma{}^{d]}-m e_\sigma{}^{[c} R_\rho{}^{d]}\right)-\frac{1}{3} m^2 R e_\rho{}^{[c} e_\sigma{}^{d]}\right].
    \end{aligned}
\end{equation}
In perfect analogy to the subsubsect. \ref{caseb}, in the two-step breaking case, considering the rescaled vierbein $\tilde{e}_\mu{}^{a}=m e_\mu{}^{a}$ and recalling that $R_{\mu \nu}^{(0) a b}=-R_{\nu \mu}^{(0) a b}$, we obtain
\begin{equation}
    \begin{aligned}
             S=\frac{a_{CG}}{4}\int d^4 x \epsilon^{\mu \nu \rho \sigma} \epsilon_{a b c d}&\left[R_{\mu \nu}^{(0) a b}-\frac{1}{2}\left(\tilde{e}_\mu{}^{[a} R_\nu{}^{b]}-\tilde{e}_\nu{}^{[a} R_\mu{}^{b]}\right)+\frac{1}{3} R \tilde{e}_\mu{}^{[a} \tilde{e}_\nu{}^{b]}\right]\\
             &\left[R_{\rho \sigma}^{(0) c d}-\frac{1}{2}\left( \tilde{e}_\rho{}^{[c} R_\sigma{}^{d]}- \tilde{e}_\sigma{}^{[c} R_\rho{}^{d]}\right)+\frac{1}{3} R \tilde{e}_\rho{}^{[c} \tilde{e}_\sigma{}^{d]}\right],
    \end{aligned}
\end{equation}
which is equal to 
\begin{align}
             S=\frac{a_{CG}}{4}\int d^4 x \epsilon^{\mu \nu \rho \sigma} \epsilon_{a b c d}C_{\mu \nu}{}^{a b}C_{\rho \sigma}{}^{c d},
\end{align}
where $C_{\mu \nu}{}^{a b}$ is the Weyl conformal tensor. This action, as discussed in \ref{caseb}, leads to the well-know four-dimensional scale invariant Weyl action,
\begin{equation}
S =2a_{CG}\int \mathrm{d}^4 x\left(R_{\mu \nu} R^{\nu \mu}-\frac{1}{3} R^2\right).
\end{equation}

\section{Unification of Conformal gravities with internal interactions
based on the gauge group \texorpdfstring{$SO(2,16)$}{SO(2,16}}\label{sec5}

We start as in the ordinary gravity from the gauge group $SO(18)$ \cite{Konitopoulos:2023wst} in the tangent space with the aim to describe the CG based on $SO(2,4) \sim SU(2,2) \sim SO(6) \sim SU(4)$ (the last two in Euclidean signature) unified with Internal Interactions based on the gauge group $SO(10)$. Therefore $C_{SO(2,16)} (SO(2,4)) = SO(12)$, which should break further to $SO(10)$, as will be discussed in the present section.  

Let us consider the breaking of $SO(18)$ to its maximal subgroup $SU(4) \times SO(12)$ and use Euclidean signature for simplicity (the non-compact case will be discussed at the end). This can be done using the $170$ rep of $SO(18)$ given that the branching rules \cite{Feger_2020} are :
\begin{equation}
\begin{aligned} 
     SO(18) &\supset SU(4) \times SO(12) \\
    18   &= (6,1) + (1,12)\qquad \;\, \quad \qquad\qquad\qquad\qquad\text{vector}\\
   153   &= (15,1) + (6,12) + (1,66)  \quad\qquad\qquad\qquad \text{adjoint}\\
   256   &= ( 4, \bar{32}) + (\bar{4}, 32)\qquad \, \quad \!\qquad\qquad\qquad\qquad \text{spinor}\\
   170   &= (1,1) + (6,12) + (20',1) + (1,77)\qquad \quad \;\text{2nd rank symmetric}\\
\end{aligned}
\end{equation}
According to the above, by giving vev in the $\langle 1,1 \rangle$ component of a scalar in the $170$ rep after SSB we obtain $SU(4) \times SO(12)$.

In order to further break the $SO(12)$ down to $SO(10) \times U(1)_\text{global}$ or to $SO(10) \times U(1)$ we can use scalars either in the $66$ rep (contained in the adjoint $153$ of $SO(18)$) or in the $77$ (contained in the $170$ of $SO(18)$) respectively, given the branching rules,
\begin{equation}
\begin{aligned} 
   SO(12) &\supset SO(10) \times U(1)\\
     66   &= (1)(0) + (10)(2) + (10)(-2) + (45)(0)\\
 77   &= (1)(4) + (1)(0) + (1)(-4) + (10)(2) + (10)(-2) + (54)(0).
\end{aligned}
\end{equation}
According to the above, by giving vev to the $\langle (1)(0)\rangle$ of the $66$ rep we obtain $SO(10) \times U(1)$ after the SSB and by giving vev to the $\langle(1)(4)\rangle$ of the $77$ rep we obtain $SO(10) \times U(1)_{\text{global}}$ after the SSB.

Similarly, we can further break $SU(4)$ down to $SO(4) \sim SU(2) \times SU(2)$ in two steps. First we break it to $SO(2,3) \sim SO(5)$, and then to $SO(4)$. For that recall the following branching rules \cite{Slansky:1981yr,Feger_2020}:
\begin{equation}
\begin{aligned} 
   SU(4) &\supset SO(5) \\
     4   &=  4\\
     6   &= 1 + 5.
\end{aligned}
\end{equation}
Then by giving vev to a scalar that belongs on the $6$ rep of $SU(4)$ (which belongs in the $18$ rep of $SO(18)$) in the $\langle 1\rangle$ component, the $SU(4)$ breaks down to the $SO(5)$\footnote{Note that $SO(5) \sim SO(2,3)$, i.e. the AdS group that has been discussed in section \ref{sec2}.}.

Then according to the branching rules:
\begin{equation}
\begin{aligned} 
  SO(5) &\supset SU(2) \times SU(2)\\
    5  &= (1,1) + (2,2)\\
    4  &= (2,1) + (1,2),
\end{aligned}
\end{equation}
by giving vev in $\langle 1,1\rangle $ of a scalar in the $5$ rep of $SO(5)$ (contained in the $6$ of $SU(4)$) and in the $18$ of $SO(18)$) we eventually obtain the Lorentz group $SU(2) \times SU(2) \sim SO(4) \sim SO(1,3)$.

It should be noted that there exist two more ways to break the $SU(4)$ gauge group to $SU(2) \times SU(2)$. The first new way to break $SU(4)$ to $SU(2) \times SU(2)$ is the direct one, using a scalar in the symmetric 2nd rank tensor rep \cite{Li:1973mq}. In that case we have:
\begin{equation}
\begin{aligned} 
  SU(4) &\supset SU(2) \times SU(2)\\
    4  &= (2,2)\\
    10  &= (1,1) + (3,3).\\
\end{aligned}
\end{equation}
Then we see that by giving vev in the $\langle 1,1 \rangle$ of the scalar in the $10$ rep, which is the symmetric 2nd rank tensor of $SU(4)$ and contained in the $816$ rep of $SO(18)$, after SSB we obtain $SU(2) \times SU(2)\sim SO(1,3)$. However, we also see that the $4$, which will accommodate the fermions decomposes under $SU(2) \times SU(2)$ as $(2,2)$ and not as $(2,1)$ and $(1,2)$ in order to be identified with the two components Weyl spinors of SO(1,3) (see discussion in section \ref{sec6}).

The other way to break the $SU(4)$ gauge group to $SU(2) \times SU(2)$ is to use scalars in the adjoint rep of $SU(4)$, $15$, which is contained in the adjoint rep
of $SO(18)$, $153$. In that case we have:
\begin{equation}
\begin{aligned} 
  SU(4) &\supset SU(2) \times SU(2)\times U(1)\\
    4   &= (2,1)(1) + (1,2)(-1)\\
    15  &= (1,1)(0) + (2,2)(2) + (2,2)(-2) + (3,1)(0) + (1,3)(0)
\end{aligned}
\end{equation}
Then by giving vev in the $\langle 1,1\rangle$ direction of the adjoint rep $15$ we obtain the known result \cite{Li:1973mq}, that $SU(4)$ breaks spontaneously to $SU(2) \times SU(2) \times U(1)$. The way to vanish the corresponding $U(1)$ gauge boson and remain with the $SU(2) \times SU(2)$ was discussed already in subsection \ref{AdjointSubsection} (see also footnote \ref{footnote}). Note in addition that in this case the $4$ is decomposed in the appropriate reps of $SU(2) \times SU(2) \sim SO(1,3)$ to describe the two Weyl spinors. Therefore, we choose the latter way to break $SU(4)$. At this point, it is also noted that the same breaking pattern could be obtained in the $SO(6)(\sim SU(4))$ using the antisymmetric 2nd rank tensor $15$:
\begin{equation}
    SO(6) \rightarrow SU(2)\times SU(2) \times U(1),
\end{equation}
as it can be seen in appendix \ref{apb}.

Having established the analysis of the various breakings using the branching rules under the maximal subgroups starting from the group $SO(18)$, one can easily consider instead the isomorphic algebras of the various groups. Specifically, instead of  $SO(18)$, the isomorphic algebra of the non-compact groups $SO(2,16) \sim SO(18)$, and similarly $SO(2,4) \sim SO(6) \sim SU(4)$ and $SO(2,3) \sim Sp_4 \sim SO(5)$, keeping in mind the SSB as it has been discussed in section \ref{sec2} and \ref{sec3}.

\section{Fermions}\label{sec6}
Having examined the various breakings in section \ref{sec5}, let us next discuss the fermions.
  
A Dirac spinor, $\psi$ has $2^{D/2}$ independent components in $D$ dimensions. Then the Weyl and Majorana conditions when imposed each divide the number of independent components by $2$. The Weyl  constraint can be imposed only for even $D$, therefore the Weyl-Majorana spinor resulting after imposing to a Dirac spinor both Weyl and Majorana conditions has $2^{(D-4)/2}$ independent components (for $D$ even).

The unitary reps of the Lorentz group $SO(1,D-1)$ are labeled by a continuous momentum vector $\mathbf{k}$, and by a spin `projection', which in $D$ dimensions is a rep of the compact subgroup $SO(D-2)$. The Dirac, Weyl, Majorana, and Weyl-Majorana spinors carry indices that transform as finite-dimensional non-unitary spinor reps of $SO(1,D-1)$.

It is also known \cite{D_Auria_2001, majoranaspinors}, that the type of spinors one obtains for $SO(p,q)$ in real case is governed by the signature $(p-q)$ $\text{mod}(8)$. Among even signatures, signature zero gives a real rep, signature four a quaternionic rep, while signatures two and six give complex reps. In the case of $SO(2,16)$ the signature is six, and imposing the Majorana condition in addition to Weyl is permitted.

For completeness and fixing the notation let us recall, the well-known case of $4$ dimensions. The $SO(1,3)$ spinors in the usual $SU(2) \times SU(2)$ basis transform as $(2,1)$ and $(1,2)$, with reps labeled by their dimensionality. The two-component Weyl spinors, $\psi_L$ and $\psi_R$, transform as the irreducible spinors, $\psi_L \sim  (2, 1)$ $\psi_R \sim (1, 2)$, of $SU(2) \sim SU(2)$ with ``$\sim$'' here meaning ``transforms as''. Then a Dirac spinor, $\psi$, is made from the direct sum of $\psi_L$ and $\psi_R$, $\psi \sim (2, 1) \oplus (1, 2)$. Accordingly in four-component notation the Weyl spinors in the Weyl basis are $(\psi_L, 0)$ and $(0, \psi_R)$, and are eigenfunctions of $\gamma^5$ with eigenvalues $-1$ and $+1$, respectively.

The usual Majorana condition for a Dirac spinor has the form, $ \psi = C \bar{\psi}^T$, where $C$ is the charge-conjugation matrix. In four dimensions $C$ is off-diagonal in the Weyl basis, since it maps the components transforming as $(2, 1)$ into $(1, 2)$.

For even $D$, it is always possible to define a Weyl basis where $\Gamma^{D+1}$ (which consists of the product of all $\Gamma$ matrices in $D$ dimensions) is diagonal, therefore
\begin{equation}\label{a}
    \Gamma^{D+1} \psi_\pm = \pm \psi_\pm.
\end{equation}
We can express $\Gamma^{D+1}$ in terms of the chirality operators in four and extra $d$ dimensions,
\begin{equation}\label{b}
    \Gamma^{D+1} = \gamma^5 \otimes \gamma^{d+1}.               
\end{equation}
As a result the eigenvalues of $\gamma^5$ and $\gamma^{d+1}$ are interrelated. It should be noted though that the choice of the eigenvalue of $\Gamma^{D+1}$ does not impose the eigenvalues on the separate $\gamma^5$ and $\gamma^{d+1}$.

Given that $\Gamma^{D+1}$ commutes with the Lorentz generators, then each of the $\psi_+$ and $\psi_-$ corresponding to its two eigenvalues, according to eq. \eqref{a}, transforms as an irreducible spinor of $SO(1,D-1)$. For $D$ even, the $SO(1,D-1)$ always has two independent irreducible spinors; for $D = 4n$ there are two self-conjugate spinors $\sigma_D$ and ${\sigma_D}^\prime$, while for $D = 4n + 2$, $\sigma_D$ is non-self-conjugate and $\bar{\sigma}_D$ is the other spinor. Conventionally is selected $\psi_{-} \sim \sigma_D$ and $\psi_+ \sim {\sigma_D}^\prime$ or $\bar{\sigma}_D$. Then, Dirac spinors are defined as direct sum of Weyl spinors,
\begin{equation}
    \psi  = \psi_+ \oplus \psi_- \thicksim 
    \begin{cases} \sigma_D \oplus {\sigma_D}^\prime \; &\text{for}\; D = 4n \\
 \sigma_D \oplus \bar{\sigma_D}\; &\text{for}\; D = 4n+2.
    \end{cases}         
\end{equation}
  
The Majorana condition can be imposed in $ D = 2, 3, 4 + 8n$ dimensions and therefore the Majorana and Weyl conditions are compatible only in $D = 4n + 2$ dimensions.

We limit ourselves here in the case that $D = 4n + 2$ ( for the rest see e.g. refs \cite{CHAPLINE1982461, KAPETANAKIS19924}). Then starting with Weyl-Majorana spinors in $D = 4n + 2$ dimensions, we are actually forcing a rep, $f_R$, of a gauge group defined in higher dimensions to be the charge conjugate of $f_L$, and we arrive in this way to a four-dimensional theory with the fermions only in the $f_L$ rep of the gauge group.

Here we start, keeping again the Euclidean signature, with the Weyl spinor of $SO(18)$, $256$ and according to the breakings and branching rules discussed in section \ref{sec5} we have
\begin{equation}
    \begin{aligned}
    SO(18) &\supset  SU(4)\times SO(12)\\
    256 &= (4,\bar{32}) + (\bar{4},32).
    \end{aligned}
\end{equation}

Given that the Majorana condition can also be imposed we are led to have fermions in the $(4, \bar{32})$ of $SU(4)\times SO(12)$.

Then we have the following branching rule of the $32$ under the $SO(10)\times [U(1)]$
\begin{equation}
\label{so12breaking}
    \begin{aligned}
    SO(12) &\supset SO(10)\times [U(1)]\\
    32 &= (\bar{16})(1) + (16)(-1) \Rightarrow \\  
    \bar{32} &= (16)(-1) + (\bar{16})(1).
    \end{aligned}
\end{equation}

The $\left[U(1)\right]$ is put to take into account the case that $U(1)$ exists as gauge symmetry and the case that it is broken (see the breaking of $SO(12)$ with scalar in the $77$ rep in eq. \eqref{so12breaking}) leaving a $U(1)$ as a remaining global symmetry.

On the other hand, as noted earlier,
\begin{equation}
    \begin{aligned}
    SU(4) &\rightarrow SU(2)\times SU(2)\\
    4 &= (2,1) + (1,2).
    \end{aligned}
\end{equation}

Therefore we obtain after all breakings:
\begin{equation}
    \begin{aligned}
    SU(2)\times SU(2)\times SO(10)\times [U(1)]\{[(2,1) + (1,2)\} \{(16)(-1) + \bar {16})(1)\}\\
    = ((16)(-1)_L+ ((\bar{16})(1))_L+ ((16)(-1))_R+ (\bar{16})(1)_R,
\end{aligned}
\end{equation}

and since $((\bar{16})(1))_R = ((16)(-1)_L$ and $((\bar{16})(1))_L = ((16)(-1)_L)$,
\begin{equation}
    =2\times (16)_L(-1) + 2\times (16)_R(-1).
\end{equation}

Finally choosing to keep only the $-1$ eigenvalue of $\gamma^5$ we obtain:
\begin{equation}
    2\times (16)_L(-1).
\end{equation}
Similarly to the general discussion we presented earlier in this section concerning the Weyl condition in $D$ and $4$ dimensions, namely that they are independent to each other, the same holds for the Majorana condition. If we impose the Majorana condition in higher dimensions we are still free to impose the Majorana condition once more in lower dimensions, taking into account the rule for the non-compact groups $SO(p,q)$ mentioned earlier in the present section. Therefore if we impose in addition the Weyl also the Majorana condition in higher dimensions we can still impose the the same conditions in lower dimensions, respecting the known rules for each case.

Therefore, given the above analysis the gauge group describing the Internal Interactions is $C_{SO(18)}(SO(6)) = SO(10)\times U(1)_{\text{global}}$, while the type of spinors that we have is governed by the signature of $(p-q)$ that permits the imposition of Weyl and Majorana conditions in higher 
and four dimensions leading to one generation of $16_L$ in $SO(10)$. Obviously the other generations are introduced as usual with more spinors in $SO(2,16)$.


\section{Conclusions}\label{sec7}
The present work is based on the observation that the dimension of the tangent space is not necessarily equal to the dimension of the corresponding curved manifold. Combining this idea together with the fact that gravitational theories can be described by gauge theories, as the usual internal interactions, we were able to provide unification of all interactions using the higher-dimensional tangent group $SO(2,16)$ in four dimensions. Therefore in this way the old dream of unifying all interactions using extra dimensions is realised, while the whole discussion is kept in four dimensions, i.e. without the need of compactification and dimensional reduction. Therefore the celebrated theoretical idea of extra dimensions remains as a very useful theoretical tool, despite the non-observation of more than four dimensions.

An obvious question concerning the unification of spacetime symmetries and internal interactions is how we evade the Coleman-Mandula (CM) theorem \cite{Coleman1967}. Recall that the CM theorem has several hypotheses with the most relevant being that the theory is Poincar\'{e} invariant. In previous studies \cite{Percacci:1984ai,Percacci_1991,Nesti_2008,Nesti_2010,Krasnov:2017epi,Chamseddine2010,Chamseddine2016,Manolakos:2023hif, noncomtomos,Konitopoulos:2023wst} the unifying group did not contain the whole Poincar\'{e} group, but rather the
Lorentz rotations. Here the Poincar\'{e} group appears spontaneously broken.

We have presented first the gauging of the AdS group $SO(2,3)$ to recall the basics of the description of gravity by gauging the tangent group. Moreover in this example we had the opportunity to recall the importance of SSB that leads eventually to GR. Based on the latter we could develop in a straightforward manner the strategy of unifying all interactions as a gauge theory of the extended four-dimensional gauge group and to apply it in the specific case of $SO(2,16)$ as the unifying group that unifies the CG based on the gauge group $SO(2,4)$, with internal interactions based on the GUT $SO(10)$. The use of SSB mechanism led to GR and WG in all cases. In a future work we plan to examine the phenomenological and cosmological implications of these constructed theories.

\section*{Acknowledgements}
It is a pleasure to thank Alex Kehagias and Pantelis Manousselis for several discussions and suggestions on the content of the paper. We would also like to thank Costas Bachas, Emilian Dudas, Dumitru Ghilencea, Anamaria Hell, Dieter L\"ust, Roberto Percacci, and Thanassis Chatzistavrakidis for reading the manuscript and their comments. D.R. would like to thank NTUA for a fellowship for doctoral studies.
G.Z. would like to thank MPP-Munich, ITP-Heidelberg and DFG Exzellenzcluster
2181:STRUCTURES of Heidelberg University for their hospitality and support. 

\section*{Appendices}
\appendix

\section{Algebras of the Anti de-Sitter and the Conformal Group}
\label{apa}
\subsection{Anti-de Sitter gauge group algebra}
\label{apa1}
The algebra of the $SO(2,3)$ group in five-dimensional notation is the following:
\begin{equation}
    \left[J_{A B}, J_{C D}\right] = \eta_{BC}J_{AD}+\eta_{AD} J_{BC} -\eta_{AC} J_{BD} -\eta_{BD} J_{AC} ,
\end{equation}
where the metric of the gauge theory is $\eta_{AB}=\operatorname{diag}(-1,1,1,1,-1)$ and $A,B=1,\dots,5$.

The decomposition of the above algebra in four-dimensional notation, in which $a,b=1,\dots,4$ yields:
\begin{equation}
\begin{aligned}
\left[J_{ab}, J_{cd}\right] &= \eta_{bc} J_{ad} +\eta_{ad} J_{bc} - \eta_{ac}J_{bd}-\eta_{bd} J_{ac},\\
\left[J_{ab}, J_{c5}\right] &= \eta_{bc} J_{a5} - \eta_{ac}J_{b5},\\
\left[J_{a5}, J_{b5}\right] &= J_{ab},
\end{aligned}
\end{equation}
where $\eta_{ab}=\operatorname{diag}(-1,1,1,1)$.

Identifying $J_{a5}$ (i.e. the set of the broken generators) with the translation generators, $P_a$ with corresponding gauge fields the vierbeine $e_{\mu}{}^{a}$ and the unbroken to the Lorentz, $M_{ab}$ with corresponding gauge fields $\omega_{\mu}{}^{a b}$, we obtain:
   \begin{equation}
    \begin{aligned}
   {\left[M_{a b}, M_{c d}\right] } & =\eta_{b c} M_{a d}+\eta_{a d} M_{b c}-\eta_{a c} M_{b d}-\eta_{b d} M_{a c},\\
\left[M_{ab}, P_{c}\right] &= \eta_{bc} P_{a} - \eta_{ac}P_{b},\\
\left[P_{a}, P_{b}\right] &= M_{ab}.
\end{aligned}
\end{equation}
The above commutation relations comprise the AdS group algebra decomposed in four-dimensional notation. 

In the case in which the breaking occurs through the SSB presented in section \ref{sec2}, the broken generators, $P_a$, are being rescaled as $m^{-1}P_a$, and the last commutation relation becomes:
 \begin{equation}
\left[P_{a}, P_{b}\right] =m^2 M_{ab}.
\end{equation}

\subsection{Conformal Group algebra}
\label{apa2}
Like in the AdS case we begin with the $SO(2,4)$ algebra in six-dimensional notation:
\begin{equation}
    \left[J_{A B}, J_{C D}\right] = \eta_{BC}J_{AD}+\eta_{AD} J_{BC} -\eta_{AC} J_{BD} -\eta_{BD} J_{AC} ,
\end{equation}
where the metric of the gauge theory is $\eta_{AB}=\operatorname{diag}(-1,1,1,1,-1,1)$ and $A,B=1,\dots,6$.

In five-dimensional notation the above algebra becomes:
  \begin{equation}
    \begin{aligned}
    \left[J_{ij}, J_{kl}\right] &= \eta_{jk}J_{il}+\eta_{il} J_{jk}-\eta_{ik} J_{jl} -\eta_{jl} J_{ik} ,\\
\left[J_{ij}, J_{k6}\right] &= \eta_{jk}J_{i6}-\eta_{ik} J_{j6},\\
\left[J_{i6}, J_{j6}\right] &= -J_{ij},
\end{aligned}
\end{equation}
where $i,j=1,\dots,5$, and $\eta_{ij}=\operatorname{diag}(-1,1,1,1,-1)$.
Performing one more step of decompositions, i.e. passing to four-dimensional notation, we obtain:
  \begin{equation}
    \begin{aligned}
    \left[J_{ab}, J_{cd}\right] &= \eta_{bc} J_{ad} +\eta_{ad} J_{bc} - \eta_{ac}J_{bd}-\eta_{bd} J_{ac},\\
\left[J_{ab}, J_{c5}\right] &= \eta_{bc} J_{a5} - \eta_{ac}J_{b5},\\
\left[J_{a5}, J_{b5}\right] &= J_{ab},\\
\left[J_{ab}, J_{c6}\right] &= \eta_{bc} J_{a6} - \eta_{ac}J_{b6},\\
\left[J_{a6}, J_{b6}\right] &= -J_{ab}\\
\left[J_{ab}, J_{56}\right] &= \left[J_{56}, J_{56}\right] =  0\\
\left[J_{a5}, J_{56}\right] &=- J_{a6}\\
\left[J_{a6}, J_{56}\right] &= -J_{a5} \\
\left[J_{a5}, J_{b6}\right] &= -\eta_{ab}J_{56}
\end{aligned}
\end{equation}
where $a,b=1,\dots,4$, and $\eta_{ab}=\operatorname{diag}(-1,1,1,1)$.

In order to identify the $J_{a5},J_{a6}$, and $J_{56}$ generators to the conformal algebra generators we have two choices: The first choice is to regard the four dimensional conformal algebra as isomorphic to the five-dimensional AdS algebra, while the second is to regard it as an extension of the Poincar\'{e} algebra. Choosing one of the two will determine whether the translations generators will be commuting or not. In the current work we chose to regard the conformal algebra as an extension of the Poincar\'{e}, and thus we set the following:
\begin{equation}
    \begin{aligned}
    M_{ab}&=J_{ab},\qquad
        P_a&=-J_{a5}-J_{a6},\qquad
        K_a&=-J_{a5}+J_{a6},\qquad
        D&=-J_{56}.        
    \end{aligned}
\end{equation}
The commutation relations we result with are the well-known:
\begin{equation}
\label{commutatorz}
\begin{aligned}
{\left[M_{a b}, M_{c d}\right] } & =\eta_{b c} M_{a d}+\eta_{a d} M_{b c}-\eta_{a c} M_{b d}-\eta_{b d} M_{a c}, \\
{\left[M_{a b}, P_c\right] } & =\eta_{b c} P_a-\eta_{a c} P_b, \\
{\left[M_{a b}, K_c\right] } & =\eta_{b c} K_a-\eta_{a c} K_b, \\
{\left[P_a, D\right] } & =P_a, \\
{\left[K_a, D\right] } & =-K_a, \\
{\left[K_a, P_b\right] } & =-2\left(\eta_{a b} D+M_{a b}\right).
\end{aligned}
\end{equation}

In this work, the representation chosen for the above generators is the one used in \cite{Kaku:1978nz} and is the following:
\begin{equation}
\begin{gathered}
\label{generatorsrep}
M_{a b}=\frac{1}{4}[\gamma_{a},\gamma_{b}], \quad K_a=\frac{1}{2}\gamma_a(1+\gamma_5),\quad P_a=-\frac{1}{2}\gamma_a(1-\gamma_5), \quad D=-\frac{1}{2}\gamma_5, 
\end{gathered}
\end{equation}
in which the rep of the gamma matrices chosen is the following:
\begin{equation}
\begin{gathered}
\{\gamma_a ,\gamma_b\}=-2\eta_{ab},\quad \text{where}\quad \eta_{ab}=\operatorname{diag}(-1,1,1,1), \quad \text{and}\quad \gamma_5=i\gamma_2 \gamma_3 \gamma_4\gamma_1, 
\end{gathered}
\end{equation}
where
\begin{equation}
\begin{aligned}
\gamma^1 & =\left(\begin{array}{cccc}
0 & 0 & 1 & 0 \\
0 & 0 & 0 & 1 \\
1 & 0 & 0 & 0 \\
0 & 1 & 0 & 0
\end{array}\right), & \gamma^2=\left(\begin{array}{cccc}
0 & 0 & 0 & 1 \\
0 & 0 & 1 & 0 \\
0 & -1 & 0 & 0 \\
-1 & 0 & 0 & 0
\end{array}\right), \\ \\
\gamma^3 & =\left(\begin{array}{cccc}
0 & 0 & 0 & -i \\
0 & 0 & i & 0 \\
0 & i & 0 & 0 \\
-i & 0 & 0 & 0
\end{array}\right), & \gamma^4=\left(\begin{array}{cccc}
0 & 0 & 1 & 0 \\
0 & 0 & 0 & -1 \\
-1 & 0 & 0 & 0 \\
0 & 1 & 0 & 0
\end{array}\right), 
\end{aligned}
\end{equation}
and 
\begin{equation}
\gamma^5 =\left(\begin{array}{cccc}
1 & 0 & 0 & 0 \\
0 & 1 & 0 & 0 \\
0 & 0 & -1 & 0 \\
0 & 0 & 0 & -1
\end{array}\right).
\end{equation}
The anticommutation relations of the generators are also required, and are the following:
\begin{equation}
\label{anticommutatorz}
\begin{aligned}
 \{M_{a b}, M_{c d}\}&=\frac{1}{2}\left(\eta_{a c}\eta_{b d}-\eta_{b c} \eta_{a d}\right)-i\epsilon_{abcd}D, \\
\{M_{a b}, P_c\} &=+i\epsilon_{abcd}P^d, \\
\{M_{a b}, K_c\}&=-i\epsilon_{abcd}K^d, \\
\{M_{ab},D\}&= 2M_{ab}D,\\
\{P_{a}, K_{b}\}&=4M_{ab}D+\eta_{ab},\\
\{K_{a}, K_{b}\}&=\{P_{a}, P_{b}\}=-\eta_{a b},\\
\{P_{a}, D\}&=\{K_{a}, D\}=0.
\end{aligned}
\end{equation}

\section{SSB of the conformal group by using a scalar in the antisymmetric 2nd
rank tensor representation}\label{apb}
It is interesting to examine also the breaking of the conformal gauge
group using a scalar in the antisymmetric rep instead of two scalars in the vector rep, as it was discussed in \ref{31}. Such a rep can
result from the tensor product of two vector reps, $6$, of $SO(6) ( \sim
SO(2,4))$.
Given that $6 \times 6 = 1_s + 15_a + 20_{s}^{\prime}$ we can choose to consider the SSB by a scalar in the $15_a$ rep of $SO(6)$. In the case of ordinary Higgs
minimization the SSB of $SO(6)$ could be either $U(3)$ or $SU(2) \times SU(2) \times U(1)$
\cite{Li:1973mq}, depending from the sign of one of coefficients of the self coupling of the scalar field. Either of the two possibilities can be selected by constraining  the form of the scalar field in the vacuum (see \cite{Li:1973mq}). Here we concentrate in the second option. In this case the action of the conformal group along with the scalar
field in the antisymmetric 2nd rank tensor rep would be
\begin{equation}
\begin{aligned}
    S_{SO(2,4)}=a_{CG}\int d^4x [&\epsilon^{\mu \nu \rho \sigma}\epsilon_{ABCDEF}\phi^{EF}\sqrt{2}m\frac{1}{4}F_{\mu \nu}{}^{AB}F_{\rho \sigma}{}^{CD}+\lambda (\phi^{EF} \phi_{EF} +{m}^{-2})], 
\end{aligned}
\end{equation}
where the scalar field is fixed in the vacuum to have the form given in
ref \cite{Li:1973mq}:
\begin{equation}
\phi^{EF}=\phi^0= \frac{m^{-1}}{\sqrt{2}}\left(\begin{array}{cccccc}
0 & 0 & 0 & 0 & 0 & 0 \\
0 & 0 & 0 & 0 & 0 & 0 \\
0 & 0 & 0 & 0 & 0 & 0 \\
0 & 0 & 0 & 0 & 0 & 0 \\
0 & 0 & 0 & 0 & 0 & 1 \\
0 & 0 & 0 & 0 & -1 & 0
\end{array}\right),
\end{equation}
and the component gauge fields $b,\tilde{a}$ and $e$ can be scaled in the same manner as presented above. If we additionally set the gauge field $\tilde{a}_\mu$, which corresponds to the $U(1)$ generator, equal to zero, for the reasons explained in \ref{31}, we result with $SU(2)\times SU(2)\sim SO(1,3)$ as the final symmetry. The resulting action in this case is identical to eq. \eqref{act2}, which can yield GR in AdS space, \eqref{34} and \eqref{35}, or Weyl's theory of gravity, \eqref{weyll}.

\printbibliography

\end{document}